\documentclass[aps,pra,twocolumn,showpacs,showkeys]{revtex4}
\usepackage{amsmath}
\begin{document}

\title{The generalized partial transposition criterion \\
for separability of multipartite quantum states}
\author{Kai Chen}
\email{kchen@aphy.iphy.ac.cn}
\author{Ling-An Wu}
\email{wula@aphy.iphy.ac.cn}
\affiliation{Laboratory of Optical Physics, Institute of Physics, Chinese Academy of
Sciences, Beijing 100080, P.R. China,}

\begin{abstract}
We present a generalized partial transposition separability criterion for the
density matrix of a multipartite quantum system. This criterion comprises as
special cases the famous \emph{Peres-Horodecki} criterion and the recent
\emph{realignment} criterion in [O. Rudolph, quant-ph/0202121] and
[K. Chen, L.A. Wu, quant-ph/0205017]. It involves
only straightforward matrix manipulations and is easy to apply. A quantitative
measure of entanglement based on this criterion is also obtained.
\end{abstract}

\pacs{03.67.-a, 03.65.Ud, 03.65.Ta}
\keywords{Separability;  Density matrix; Multipartite quantum system;
Measure of entanglement}
\maketitle

\date{Revised Oct 31, 2002}

\section{Introduction}

Since the well-known papers of Einstein, Podolsky and Rosen \cite{EPR35},
Schr\"{o}dinger \cite{Sch35} and Bell \cite{Bell64}, quantum entangled states
have greatly enriched quantum mechanics and have recently found wide
applications in the rapidly expanding field of quantum information processing.
Quantum teleportation, quantum cryptography, quantum dense coding and parallel
computation \cite{pre98,nielsen,zeilinger} have spurred a flurry of activity
in the effort to fully exploit the potential of quantum entanglement. Although
their applications have already been demonstrated experimentally, the physical
character and mathematical structure of entangled states are only partially
known, and a full comprehensive understanding is still a challenge for the theorists.

The most familiar entangled state is the singlet pure state of a pair of
spin-$\frac{1}{2}$ particles given by Bohm \cite{Bohm}
\begin{equation}
\psi_{s}={\frac{1}{\sqrt{2}}}(|\uparrow\rangle|\downarrow\rangle
-|\downarrow\rangle|\uparrow\rangle),
\end{equation}
which cannot be reduced or factorized to a direct product of the states for
the two particles. Due to uncontrolled interactions with the environment, a
practical composite system generally evolves to a mixed state. How do we know
if a quantum state is entangled, and how entangled is it still after the
intervention of noise?

To answer these questions we must have a physical acceptable definition of
entangled states. From a physically meaningful and practical point of view,
the state of a composite quantum system is called \emph{disentangled} or
\emph{separable} if it can be prepared in a ``\emph{local}"
or ``\emph{classical}"
way. A separable multipartite system can be expressed as an ensemble
realization of pure product states $\left\vert \psi_{i}\right\rangle
_{A}\left\vert \phi_{i}\right\rangle _{B}\cdots\left\vert \varphi
_{i}\right\rangle _{Z}$ occurring with a certain probability $p_{i}$:
\begin{equation}
\rho_{AB\cdots Z}=\sum_{i}p_{i}\rho_{i}^{A}\otimes\rho_{i}^{B}\otimes
\cdots\otimes\rho_{i}^{Z}, \label{sep}
\end{equation}
where $\rho_{i}^{A}=\left\vert \psi_{i}\right\rangle _{A}\left\langle \psi
_{i}\right\vert $, $\rho_{i}^{B}=\left\vert \phi_{i}\right\rangle
_{B}\left\langle \phi_{i}\right\vert $, $\cdots$, $\rho_{i}^{Z}=\left\vert
\varphi_{i}\right\rangle _{Z}\left\langle \varphi_{i}\right\vert $, $\sum
_{i}p_{i}=1$, and $\left\vert \psi_{i}\right\rangle _{A}$, $\left\vert
\phi_{i}\right\rangle _{B}$, $\cdots$, $\left\vert \varphi_{i}\right\rangle
_{Z}$ are normalized pure states of subsystems $A$,$B$,$\cdots$, and $Z$,
respectively\cite{werner89}. If no convex linear combination exists for a
given $\rho_{AB\cdots Z}$, then the state is called ``\emph{entangled}"
and includes quantum correlation.

For a pure state $\rho_{AB\cdots Z}$, it is trivial and straightforward to
judge its separability: \vskip0.2cm \noindent\emph{A pure state $\rho
_{AB\cdots Z}$ is separable if and only if
\begin{equation}
\rho_{AB\cdots Z}=\rho_{A}\otimes\rho_{B}\otimes\cdots\otimes\rho_{Z},
\end{equation}
where $\rho_{A,B,\cdots,Z}$ are the reduced density matrices defined as
$\rho_{A}=Tr_{B,C,\cdots,Z}(\rho_{AB\cdots Z})$, $\rho_{B}=Tr_{A,C,\cdots
,Z}(\rho_{AB\cdots Z})$, $\cdots,\rho_{Z}=Tr_{A,B,\cdots,Y}(\rho_{AB\cdots
Z})$.} \vskip0.2cm

\noindent However, for a generic mixed state $\rho_{AB\cdots Z}$, finding a
decomposition as in Eq.~(\ref{sep}) or proving that it does not exist is a
non-trivial task (we refer to recent reviews \cite{lbck00,terhal01,3hreview}
and references therein). There has been considerable effort in recent years to
analyze the separability and quantitative character of quantum entanglement.
The Bell inequalities satisfied by a separable system give the first necessary
condition for separability \cite{Bell64}. In 1996, Peres made an important
step forward and showed that partial transpositions with respect to one and
more subsystems of the density matrix for a separable state are positive,
\begin{equation}
\rho^{T_{\mathcal{X}}}\geq0,
\end{equation}
where $\mathcal{X}\subset\{A,B,\cdots,Z\}$. Thus the $\rho^{T_{\mathcal{X}}}$
should have non-negative eigenvalues (this is known as the $PPT$ criterion or
\emph{Peres-Horodecki} criterion) \cite{peres}. This was immediately shown by
Horodecki \textit{et al} \cite{3hPLA223} to be sufficient for bipartite
systems of $2\times2$ and $2\times3$. Meanwhile, they also found a necessary
and sufficient condition for separability by establishing a close connection
between positive map theory and separability \cite{3hPLA223}. In view of the
quantitative character for entanglement, Wootters succeeded in computing the
``\textit{entanglement of formation}" \cite{be96} and thus obtained
a separability criterion for $2\times2$
mixtures \cite{wo98}. The ``\textit{reduction criterion}" proposed
independently in \cite{2hPRA99} and \cite{cag99}
gives another necessary criterion which is equivalent to the $PPT$ criterion
for $2\times n$ composite systems but is generally weaker. Pittenger
\textit{et al} gave also a sufficient criterion for separability connected
with the Fourier representations of density matrices \cite{Rubin00}. Later,
Nielsen \textit{et al} \cite{nielson01} presented another necessary criterion
called the \textit{majorization criterion}: the decreasingly ordered vector of
the eigenvalues for $\rho_{AB}$ is majorized by that of $\rho_{A}$ or
$\rho_{B}$ alone for a separable state. A new method of constructing
\textit{entanglement witnesses} for detecting entanglement was given in
\cite{3hPLA223} and \cite{ter00,lkch00}. There are also some necessary and
sufficient criteria of separability for low rank cases of the density matrix,
as shown in \cite{hlvc00,afg01}. In addition, it was shown in \cite{wu00} and
\cite{pxchen01} that a necessary and sufficient separability criterion is also
equivalent to certain sets of equations.

However, despite these advances, a practical and convenient computable
criterion for generic bipartite systems is mainly limited to the $PPT$,
reduction and majorization criteria, as well as a recent extension of the
$PPT$ criterion based on semidefinite programs \cite{dps}. Very recently
Rudolph \cite{ru02} and the authors \cite{Chen02} proposed a new operational
criterion for separability: the \emph{realignment} criterion (named thus
following the suggestion of \cite{Horo02} which is also called the
\emph{computational cross norm} criterion given in Ref. \cite{ru02}). The
criterion is very simple to apply and shows dramatic ability to detect most of
the bound entangled states \cite{Chen02} and even genuinely tripartite
entanglement \cite{Horo02}. This is a surprising new result, for the bound
entangled states \cite{hPLA97} are ``weakly"
\ inseparable and in the past it was very hard to establish with certainty
their inseparability \cite{cag99}. Soon after, Horodecki \textit{et al }showed
that the $PPT$ criterion and realignment criterion can be equivalent to any
permutation of the indices of the density matrix \cite{Horo02}.

In this paper we generalize our realignment criterion to multipartite quantum
systems in arbitrary dimensions; then, from a different aspect, we give a
systematic construction for the \emph{generalized partial transposition}
criterion. The constructions are given in Section \ref{sec2} where the strong
$PPT$ and realignment criteria are shown as two special cases of the
criterion. Only involving straightforward matrix manipulations, it is also
very easy to apply. A quantitative measure of entanglement based on the
criterion for detecting entanglement is also obtained in Section \ref{sec3}. A
brief summary and some discussions are given in the last section.

\section{The criteria for separability}

\label{sec2}

In this section we will give two criteria for separability of the density
matrix. We first introduce the multipartite generalization of the realignment
criterion, then present the generalized partial transposition criterion as a
further generalization of the $PPT$ and realignment criteria.

\subsection{Some notation}

\label{sec2.1} The various matrix operations we shall use can be found in
\cite{hornt1,hornt}, with the following notation:

\noindent\textbf{Definition:} \emph{For each $m\times n$ matrix $A=[a_{ij}]$,
where $a_{ij}$ is the matrix entry of A, we define the vector $vec(A)$ as}
\[
vec(A)=[a_{11},\cdots,a_{m1},a_{12},\cdots,a_{m2},\cdots,a_{1n},\cdots
,a_{mn}]^{T}.
\]
Let $Z$ be an $m\times m$ block matrix with block size $n\times n$. We define
a ``realignment'' operation $\mathcal{R}$ to change $Z$ to a realigned matrix
$\widetilde{Z}$ of size $m^{2}\times n^{2}$ that contains the same elements as
$Z$ but in different positions as
\begin{equation}
\mathcal{R}(Z)\equiv\widetilde{Z}\equiv\left[
\begin{array}
[c]{c}
vec(Z_{1,1})^{T}\\
\vdots\\
vec(Z_{m,1})^{T}\\
\vdots\\
vec(Z_{1,m})^{T}\\
\vdots\\
vec(Z_{m,m})^{T}
\end{array}
\right]  .
\end{equation}

For example, a $2\times2$ bipartite density matrix $\rho$ can be transformed
as:
\begin{align}
\rho &  =\left(
\begin{array}
[c]{cc|cc}
\rho_{11} & \rho_{12} & \rho_{13} & \rho_{14}\\
\rho_{21} & \rho_{22} & \rho_{23} & \rho_{24}\\\cline{1-4}
\rho_{31} & \rho_{32} & \rho_{33} & \rho_{34}\\
\rho_{41} & \rho_{42} & \rho_{43} & \rho_{44}
\end{array}
\right) \nonumber\\
&  \longrightarrow\mathcal{R}(\rho)=\left(
\begin{array}
[c]{cccc}
\rho_{11} & \rho_{21} & \rho_{12} & \rho_{22}\\\cline{1-4}
\rho_{31} & \rho_{41} & \rho_{32} & \rho_{42}\\\cline{1-4}
\rho_{13} & \rho_{23} & \rho_{14} & \rho_{24}\\\cline{1-4}
\rho_{33} & \rho_{43} & \rho_{34} & \rho_{44}
\end{array}
\right)  .
\end{align}

\subsection{The realignment criterion}

\label{sec2.2}

Motivated by the Kronecker product approximation technique for a matrix
\cite{loan,pits}, we developed a very simple method to obtain the realignment
criterion in \cite{Chen02}.

\vskip0.2cm \noindent\textbf{The realignment criterion}: \emph{If an $m\times
n$ bipartite density matrix $\rho_{AB}$ is separable, then for the
$m^{2}\times n^{2}$matrix $\mathcal{R}(\rho_{AB})$ the trace norm
$||\mathcal{\ R}(\rho_{AB})||\equiv\sum_{i=1}^{q}\sigma_{i}(
\mathcal{R}(\rho_{AB}))$, which is the sum of all the singular values of
$\mathcal{R}(\rho_{AB})$, should be $\leq1,$ or equivalently $\log||\mathcal{R(}\rho
_{AB})||\leq0$ where $q=\min(m^{2},n^{2})$.}

\vskip0.2cm \noindent This criterion is strong enough to detect most of the
bound entangled states in the literature, as shown in \cite{Chen02}.

For multipartite systems, we have a natural generalization that was also
partially introduced in \cite{Horo02}:

\vskip0.2cm \noindent\textbf{The multipartite realignment criterion:}
\emph{\ If an $n-$partite density matrix $\rho$ is separable, then
\begin{equation}
||(\mathcal{R}_{(k)}\otimes I_{(n-k)})\rho||\leq1,\text{ \ \ \ \ \ }
k=2,3,\cdots,n\label{realignment}
\end{equation}
where the subscript indices mean that we act by $\mathcal{R}_{(k)}$ on $k$
chosen subsystems, while leaving untouched the remaining $n-k$ subsystems.} \vskip0.2cm

\noindent Here $\mathcal{R}_{(k)}$ can be a realigned matrix according to all
the possible bipartite cuts for the $k$ subsystems, in addition to \emph{all
their combinations}. It is surprisingly strong enough to detect the genuinely
tripartite bound entangled state \cite{Horo02} which is bi-separable with
respect to any bipartite cuts for the $3$ subsystems. The operation
$(\mathcal{R}_{(2)}\otimes I_{(n-2)})$ is also shown to be equivalent to
certain permutations of the indices of the density matrix \cite{Horo02}.

\subsection{The generalized partial transposition criterion}

\label{sec2.3}

We will now derive the main result of this paper:\ a generalized partial
transposition criterion for separability of multipartite quantum systems in
arbitrary dimensions.

\subsubsection{The main theorem}

Since the density matrix $\rho_{p}$ for a $d$-dimensional pure separable state
is a self-adjoint Hermitian $d\times d$ matrix with only one eigenvalue $1$,
it is evident that $\rho_{p}$ is of rank $1$ and we have naturally
\begin{equation}
\rho_{p}=u\otimes u^{\dagger}=u^{\dagger}\otimes u, \label{pure}
\end{equation}
where $u$ is a $d\times1$ \emph{column} vector satisfying $u^{\dagger}
u\equiv1$. This is possible since $u$ can be chosen to be the normalized
eigenvector of $\rho_{p}$. For the convenience of later use we define two new
operators $\mathcal{T}_{r}$, $\mathcal{T}_{c}$ and their multiplication:
\begin{align}
\mathcal{T}_{r}  &  :A\longrightarrow\text{row transposition of }A\nonumber\\
&  \Longleftrightarrow A\longrightarrow(vec(A))^{t},\label{p1}\\
\mathcal{T}_{c}  &  :A\longrightarrow\text{column transposition of
}A\nonumber\\
&  \Longleftrightarrow A\longrightarrow vec(A),\label{p2}\\
\mathcal{T}_{c}\mathcal{T}_{r}\text{ or }\mathcal{T}_{r}\mathcal{T}_{c}  &
:A\longrightarrow A^{t}, \label{p3}
\end{align}
where $t$ represents standard transposition operation. For example, a
$2\times2$ matrix $A$ can be transformed to

\begin{align*}
A &  =\left(
\begin{array}
[c]{cc}
a_{11} & a_{12}\\
a_{21} & a_{22}
\end{array}
\right)  \\
&  \longrightarrow\mathcal{T}_{r}(A)=\left(
\begin{array}
[c]{cc|cc}
a_{11} & a_{21} & a_{12} & a_{22}
\end{array}
\right)  ,\\
&  \longrightarrow\mathcal{T}_{c}(A)=\left(
\begin{array}
[c]{c}
a_{11}\\
a_{21}\\\hline
a_{12}\\
a_{22}
\end{array}
\right)  .
\end{align*}
As for a generic density matrix we have $\rho=
{\displaystyle\sum\limits_{i,j}}
\rho_{ij}\left\vert i\right\rangle \left\langle j\right\vert =
{\displaystyle\sum\limits_{i,j}}
\rho_{ij}\left\vert i\right\rangle \otimes\left\langle j\right\vert =
{\displaystyle\sum\limits_{i,j}}
\rho_{ij}\left\langle j\right\vert \otimes\left\vert i\right\rangle $ where
$\left\vert i\right\rangle ,\left\vert j\right\rangle $ are suitably selected
normalized orthogonal bases and $\left\langle i\right\vert ,\left\langle
j\right\vert $ are the corresponding transpositions, respectively (here,
for simplicity of notation, $\left\langle i\right\vert ,\left\langle j\right\vert $
are not regarded as the corresponding conjugate transpositions). Thus the
operations of $\mathcal{T}_{r}$ and $\mathcal{T}_{c}$ can be realized
conveniently:
\begin{align}
\rho\overset{\mathcal{T}_{r}}{\longrightarrow}
{\displaystyle\sum\limits_{i,j}}
\rho_{ij}\left\langle j\right\vert \otimes\left\langle i\right\vert
\overset{\mathcal{T}_{c}}{\longrightarrow}
{\displaystyle\sum\limits_{i,j}}
\rho_{ij}\left\vert j\right\rangle \otimes\left\langle i\right\vert  &
=\rho^{t},\label{p11}\\
\rho\overset{\mathcal{T}_{c}}{\longrightarrow}
{\displaystyle\sum\limits_{i,j}}
\rho_{ij}\left\vert j\right\rangle \otimes\left\vert i\right\rangle
\overset{\mathcal{T}_{r}}{\longrightarrow}
{\displaystyle\sum\limits_{i,j}}
\rho_{ij}\left\vert j\right\rangle \otimes\left\langle i\right\vert  &
=\rho^{t}.\label{p33}
\end{align}

Now we arrive at the following separability criterion for a multipartite system:

\vskip0.2cm \noindent\textbf{Theorem 1:} \emph{If an $n-$partite $d_{1}\times
d_{2}\times\cdots\times d_{n}$ density matrix $\rho$ is separable, then the
generalized partial transpositions of $\rho$ satisfy
\begin{equation}
||\rho^{\mathcal{T}_{\mathcal{Y}}}||\leq1,\text{ \ \ \ \ \ }\forall
\mathcal{Y}\subset\underset{2n}{\{\underbrace{r_{A},c_{A},r_{B},c_{B},
\cdots,r_{Z},c_{Z}}\}}\label{maintheorem}
\end{equation}
where $\mathcal{T}_{r_{k}}$ or $\mathcal{T}_{c_{k}}$ ($k=A,B,\cdots,Z$) means
transposition with respect to the row or column for the $k$th subsystem. The
superscript indices $\mathcal{T}$$_{\mathcal{Y}}$ represent partial
transpositions of every element of set $\mathcal{Y}$ on chosen subsystems,
while leaving untouched the remaining subsystems.}

\vskip0.2cm \noindent\textbf{Proof: }Applying Eq.~(\ref{sep}) for separable
states and the property Eq.~(\ref{pure}) for pure states, we have
\begin{equation}
\rho=\sum_{i}p_{i}u_{i}^{A}\otimes u_{i}^{A\dagger}\otimes u_{i}^{B}\otimes
u_{i}^{B\dagger}\otimes\cdots\otimes u_{i}^{Z}\otimes u_{i}^{Z\dagger}.
\end{equation}
The transformation of $\rho^{\mathcal{T}_{\mathcal{Y}}}$ is to make partial
transpositions of the rows or columns corresponding to some subsystems.
Without loss of generality we suppose that we only make a row transposition to
the $A$ subsystem:
\begin{equation}
\rho^{\mathcal{T}_{\{r_{A}\}}}=\sum_{i}p_{i}(u_{i}^{A})^{\dagger}\otimes
(u_{i}^{A})^{t}\otimes u_{i}^{B}\otimes u_{i}^{B\dagger}\otimes\cdots\otimes
u_{i}^{Z}\otimes u_{i}^{Z\dagger}.
\end{equation}
We shall need the following property (see Chapters $3,4$ of \cite{hornt}) of
the Kronecker product of the $m\times m$ matrix $A$ and the $n\times n$ matrix
$B$ (Property 1), as well as that of the trace norm of two matrices $A$ and $B$
of the same size (Property 2):

\begin{description}
\item \textbf{Property 1:} the non-zero singular values of $A\otimes B$ are
the positive numbers $\{\sigma_{i}(A)\sigma_{j}(B)\}$ where $\sigma_{i}(A)$
and $\sigma_{j}(B)$ are non zero singular values of $A$ and $B$ arranged in
decreasing order, respectively,
\end{description}

\begin{description}
\item \textbf{Property 2:} $||A+B||\leq||A||+||B||$,
\end{description}

\noindent so that
\begin{align}
&  ||\rho^{\mathcal{T}_{\{r_{A}\}}}||\nonumber\\
&  \leq\sum_{i}p_{i}||(u_{i}^{A})^{\dagger}\otimes(u_{i}^{A})^{t}\otimes
u_{i}^{B}\otimes u_{i}^{B\dagger}\otimes\cdots\otimes u_{i}^{Z}\otimes
u_{i}^{Z\dagger}||.\nonumber\\
&
\end{align}
It is straightforward that $u_{i}^{A},(u_{i}^{A})^{\dagger},(u_{i}^{A})^{t}$
and $(u_{i}^{A})^{\ast}$ have only one singular value $1$, respectively, due
to the normalization condition ${u_{i}^{A}}^\dagger u_{i}^{A}=1$. The same holds
for the case of $A\ $replaced with $B,C,\cdots$, or $Z$. Thus we obtain
$||(u_{i}^{A})^{\dagger}\otimes(u_{i}^{A})^{t}\otimes u_{i}^{B}\otimes
u_{i}^{B\dagger}\otimes\cdots\otimes u_{i}^{Z}\otimes u_{i}^{Z\dagger}||=1$ by
applying Property $1$. Therefore it is evident that

\begin{equation}
||\rho^{\mathcal{T}_{\{r_{A}\}}}||\leq\sum_{i}p_{i}=1.
\end{equation}
When applying column transposition, the standard transposition with respect to
one subsystem or even a combination of transpositions to a subset
$\mathcal{Y}$ of $\{r_{A},c_{A},r_{B},c_{B},\cdots,r_{Z},c_{Z}\}$, we follow a
similar procedure and obtain the same conclusion as the above-derived. Thus we
obtain the final result of Eq.~(\ref{maintheorem}). \hfill\rule{1ex}{1ex}

\subsubsection{Relationship with other necessary criteria}

We shall now show that Theorem $1$ actually encompasses previous strong
computational criteria for separability.

\paragraph{The $PPT$ criterion}

For partial transposition with respect to one or more subsystems of a
separable state, we have
\begin{equation}
\rho^{T_{\mathcal{X}}}\geq0,
\end{equation}
and $||\rho^{T_{\mathcal{X}}}||=Tr(\rho^{T_{\mathcal{X}}})=1$ due to the
Hermitian property of $\rho^{T_{\mathcal{X}}}$. It is obvious for any
$\mathcal{X}\subset\{A,B,\cdots,Z\}$ that there is a $\mathcal{Y}$ in the
generalized partial transposition criterion satisfying
\begin{equation}
||\rho^{\mathcal{T}_{\mathcal{Y}}}||=1.
\end{equation}
We only need to substitute $\mathcal{X}$ with its pairwise correspondence to
$\mathcal{Y}\subset\{r_{A}, c_{A},$ $r_{B}, c_{B}, \cdots, r_{Z}, c_{Z}\}$.
For example,
\begin{align*}
\mathcal{X}  &  =\{B,C\}\\
&  \Updownarrow\\
\mathcal{Y}  &  =\{r_{B},c_{B},r_{C},c_{C}\}.
\end{align*}

\paragraph{The multipartite realignment criterion}

\label{2.3.2.b}

From the property given in Ref. \cite{pits}, we have a Kronecker product
decomposition for a density matrix of $\rho=\sum_{i}\alpha_{i}^{A}\otimes
\beta_{i}^{B}$ where $\alpha_{i}^{A},\beta_{i}^{B}$ are, in general, not
density matrices. Thus one finds
\begin{equation}
\mathcal{R}(\rho)=\sum_{i}vec(\alpha_{i}^{A})vec(\beta_{i}^{B})^{T},
\end{equation}
it is easy to find that the above operation $\mathcal{R}(\rho)$ corresponds to
$\mathcal{Y}=\{c_{A},r_{B}\}$ from the properties of Eq.~(\ref{p1}) and
(\ref{p2}). For the generic realignment operation $(\mathcal{R}_{(k)}\otimes
I_{(n-k)})$, we will have a similar correspondence so that Theorem $1$ also
includes the multipartite realignment criterion of Eq.~(\ref{realignment})
as a special case.

\paragraph{The Horodecki indices permutation criterion}

As for the criterion proposed in \cite{Horo02}, Horodecki \textit{et al }show
that the $PPT$ criterion and the realignment criterion are equivalent to
certain permutations of density matrix indices up to some unitary matrix which
keeps the trace norm invariant. Moreover, it is obvious that the realignment
criterion is identical to the operation of $\mathcal{R}_{(2)}\otimes
I_{(n-2)}$ in the language of generic realignment, so it is also a special
case of Theorem $1$ according to (\ref{2.3.2.b}).

\vskip0.3cm

We can see that the generalized partial transposition criterion is a powerful
computational criterion since it includes as special cases the strong $PPT$
and the generic realignment criterion. This gives us a whole framework to
detect entanglement with great convenience of manipulation. In addition, it
also contains some new criteria for recognizing entanglement. For instance,
making only the row transpositions or column transpositions of various
subsystems is different from the $PPT$ and realignment criteria. However,
despite these virtues, it is still not sufficient for detecting all the
entangled states. For example, it still fails to recognize the $2\times4$
Horodecki bound entangled state \cite{hPLA97} by direct computation.

\section{Entanglement measure induced by the separability criterion}

\label{sec3}

With the power of our new criterion, we expect that it should be able to give
some information on the degree of entanglement. Among the quantitative
measures, the entanglement of formation $E_{F}(\rho)$\ \cite{be96} and
distillable entanglement $E_{D}(\rho)$ \cite{be96} are two of the most
meaningful ones. The former quantifies the asymptotic pure-state entanglement
required to create $\rho$ while the latter quantifies which can be extracted
from $\rho,$ by means of local operations and classical communication
($LOCC$). But, in practice it is very difficult to compute the two measures
for a generic quantum state. There is only one exception for the success of
computation of $E_{F}(\rho)$ for two-qubits \cite{wo98}.

Recently, Vidal and Werner showed that the negativity is an entanglement
monotone \cite{vidal98} and therefore a good entanglement measure \cite{vw01}.
The negativity is defined as
\begin{equation}
\mathcal{N}(\rho)\equiv\frac{||\rho^{T_{A}}||-1}{2}.
\end{equation}
Following this idea we introduce a computational measure based on our
criterion as
\begin{equation}
E(\rho)\equiv\sup\frac{||\rho^{\mathcal{T}_{\mathcal{Y}}}||-1}{2},\text{
\ }\mathcal{Y}\subset\{r_{A},c_{A},r_{B},c_{B},\cdots,r_{Z},c_{Z}\}
\label{measure}
\end{equation}
For separable states, $||\rho^{\mathcal{T}_{\mathcal{Y}}}||\leq1$ due to
Theorem $1$. Whenever $\mathcal{Y}$ corresponds to a standard partial
transposition of one or more of the subsystems, we have $||\rho^{\mathcal{T}
_{\mathcal{Y}}}||=1$ due to the positivity of $\rho^{\mathcal{T}_{\mathcal{Y}
}}$. Thus we have $E(\rho)=0$ for all the separable states. On the other hand,
$E(\rho)$ is convex due to the convexity of the trace norm. Furthermore, we
have the following result:

\vskip0.2cm \noindent\textbf{Theorem 2:} \emph{Applying a local unitary
transformation leaves $E(\rho)$ invariant, i.e.
\begin{equation}
E(\rho^{^{\prime}})=E(\rho),\label{equa}
\end{equation}
where $\rho^{^{\prime}}=(U_{A}\otimes U_{B}\otimes\cdots\otimes U_{Z}
)\rho(U_{A}^{\dagger}\otimes U_{B}^{\dagger}\otimes\cdots\otimes
U_{Z}^{\dagger})$ and $U_{A},U_{B},\cdots,U_{Z}$ are unitary operators acting
on A, B, $\cdots$, $Z$ subsystems, respectively.}

\vskip0.2cm \noindent\textbf{Proof:} If $||\rho^{\mathcal{T}_{\mathcal{Y}}
}||=||(\rho^{^{\prime}})^{T_{\mathcal{Y}}}||$ for $\forall\mathcal{Y}$, we
certainly have Eq.~(\ref{equa}). Consider one term of $\rho^{^{\prime}}$:
\begin{align}
\rho_{i}^{^{\prime}}  &  =(U_{A}\otimes U_{B}\otimes\cdots\otimes
U_{Z})(\alpha_{i}^{A}\otimes\beta_{i}^{B}\otimes\cdots\otimes\eta_{i}
^{Z})\nonumber\\
&  \times(U_{A}^{\dagger}\otimes U_{B}^{\dagger}\otimes\cdots\otimes
U_{Z}^{\dagger})\nonumber\\
&  =U_{A}\alpha_{i}^{A}U_{A}^{\dagger}\otimes U_{B}\beta_{i}^{B}U_{B}
^{\dagger}\otimes\cdots\otimes U_{Z}\eta_{i}^{Z}U_{Z}^{\dagger},
\end{align}
where we suppose $\rho=\sum_{i}\alpha_{i}^{A}\otimes\beta_{i}^{B}\otimes
\cdots\otimes\eta_{i}^{Z}$ which can be obtained according to \cite{lmv00} or
can be done by applying repeatedly the Kronecker product decomposition for a
bipartite-cut of a matrix \cite{loan,pits}. It should be remarked that
$\alpha_{i}^{A},\beta_{i}^{B},\cdots,\eta_{i}^{Z}$ are, in general, not
density matrices. Without loss of generality, we can just perform a row
transposition on the $A$ subsystem, thus
\begin{equation}
(\rho_{i}^{^{\prime}})^{\mathcal{T}_{\{r_{A}\}}}=(vec(U_{A}\alpha_{i}^{A}
U_{A}^{\dagger}))^{t}\otimes U_{B}\beta_{i}^{B}U_{B}^{\dagger}\otimes
\cdots\otimes U_{Z}\eta_{i}^{Z}U_{Z}^{\dagger}.
\end{equation}
Applying the property
\begin{equation}
vec(XYZ)=(Z^{T}\otimes X)vec(Y),\text{ \ \ \ \ \ \ \ \ (see \cite{hornt})}
\end{equation}
we arrive at
\begin{align}
(\rho_{i}^{^{\prime}})^{\mathcal{T}_{\{r_{A}\}}}  &  =((U_{A}^{\ast}\otimes
U_{A})vec(\alpha_{i}^{A}))^{t}\otimes U_{B}\beta_{i}^{B}U_{B}^{\dagger
}\nonumber\\
&  \otimes\cdots\otimes U_{Z}\eta_{i}^{Z}U_{Z}^{\dagger}\nonumber\\
&  =(Id\otimes U_{B}\otimes\cdots\otimes U_{Z})\nonumber\\
&  \times\left(  (vec(\alpha_{i}^{A}))^{t}\otimes\beta_{i}^{B}\otimes
\cdots\otimes\eta_{i}^{Z}\right) \nonumber\\
&  \times(U_{A}^{\dagger}\otimes U_{A}^{t}\otimes U_{B}^{\dagger}\otimes
\cdots\otimes U_{Z}^{\dagger}).
\end{align}
Since a local unitary transformation acts as a whole factor on $\rho$, we
have
\begin{align}
||(\rho^{^{\prime}})^{\mathcal{T}_{\{r_{A}\}}}||  &  =||(Id\otimes
U_{B}\otimes\cdots\otimes U_{Z})\nonumber\\
&  \times\sum_{i}\left(  (vec(\alpha_{i}^{A}))^{t}\otimes\beta_{i}^{B}
\otimes\cdots\otimes\eta_{i}^{Z}\right) \nonumber\\
&  \times(U_{A}^{\dagger}\otimes U_{A}^{t}\otimes U_{B}^{\dagger}\otimes
\cdots\otimes U_{Z}^{\dagger})||\nonumber\\
&  =||\sum_{i}\left(  (vec(\alpha_{i}^{A}))^{t}\otimes\beta_{i}^{B}
\otimes\cdots\otimes\eta_{i}^{Z}\right)  ||\nonumber\\
&  =||(\rho)^{\mathcal{T}_{\{r_{A}\}}}||.
\end{align}
Here we have used the unitarily invariant property of the trace norm.

The same procedure can be used to perform column transposition, partial
transposition of some subsystems, and any combinations of these transpositions.
The proof is thus completed. \hfill\rule{1ex}{1ex}

Since the negativity gives an upper bound for the distillable entanglement, we
also obtain an upper bound $E(\rho)$ for the distillable entanglement
$E_{D}(\rho)$, because $E(\rho)$ takes the maximum for all the possible
$\mathcal{Y}$ transpositions which naturally contain the partial transposition operations.

\section{Conclusion}

\label{sec4}

Summarizing, we have presented a computational necessary criterion for
separability of multipartite quantum systems in arbitrary dimensions which we
call a ``generalized partial transposition
criterion". This criterion unifies the previous known
\emph{PPT} criterion, the recently found realignment criterion and the
permutation criterion in a single concise framework. It provides a very
powerful necessary condition for separability and is quite easy to apply. A
quantitative measure of entanglement based on the criterion is also obtained.
Moreover, the measure gives an upper bound for distillable entanglement.

Comparing with previous works, we have significantly expanded our ability to
distinguish directly the entanglement and separability of any multipartite
quantum state in arbitrary dimensions. We expect that our method can shed some
light on the final solution of the separability problem and provide a more
suitable better measure for the degree of entanglement.

\section{Acknowledgments}

We would like to thank Dr. H. Fan and Dr. L. Yang for valuable discussions.
K.C. is grateful to encouragement from Prof. Guozhen Yang. This work was
supported by the National ``973" Program for Fundamental Research, the
Knowledge Innovation Program of the Chinese Academy of Sciences and the
National Natural Science Foundation of China.

\end{document}